\documentclass[12pt,a4paper]{article}
\usepackage[margin=1in]{geometry}
\usepackage[T1]{fontenc}
\usepackage{lmodern}
\usepackage[round,authoryear]{natbib}
\usepackage{titlesec}
\titleformat{\section}{\normalfont\Large\bfseries\raggedright}{\thesection}{1em}{}

\usepackage{amsmath,amssymb,amsthm,mathtools}
\usepackage{booktabs}
\usepackage{enumitem}
\usepackage{placeins,needspace}
\usepackage[dvipsnames]{xcolor}
\usepackage{tikz}
\usepackage{pgfplots}
\pgfplotsset{compat=1.16}
\usepgfplotslibrary{fillbetween}
\usetikzlibrary{arrows.meta,positioning,decorations.pathreplacing,calc}
\usepackage[colorlinks=true,linkcolor=MidnightBlue,citecolor=MidnightBlue,
            urlcolor=MidnightBlue]{hyperref}

\setlist{nosep}
\newtheorem{theorem}{Theorem}
\newtheorem{proposition}[theorem]{Proposition}
\newtheorem{lemma}[theorem]{Lemma}
\newtheorem{corollary}[theorem]{Corollary}
\theoremstyle{definition}
\newtheorem{assumption}[theorem]{Assumption}

\theoremstyle{remark}

\numberwithin{theorem}{section}

\newcommand{\1}{\mathbf{1}}
\newcommand{\E}{\mathbb{E}}
\newcommand{\Prb}{\mathbb{P}}

\newcommand{\proj}{\operatorname{Proj}}

\newcommand{\PhiN}{\Phi_{\!N}}

\hypersetup{
  pdftitle={Contrarian Incentives and Costly Social Learning},
  pdfauthor={Vasilii Ivanik and Georgy Lukyanov}
}

\title{Contrarian Incentives and Costly Social Learning\thanks{We are grateful to Alexis Belianin, Emiliano Catonini, Darina Cheredina, Markus Gebauer, Olivier Gossner, Vitalijs Jascisens, Margarita Kirneva, Yukio Koriyama, Fabian Slonimczyk, and Alexey Verenikin for helpful comments and discussions. We also thank participants of the ICEF research seminar. Financial support from the French National Research Agency under the \emph{Investissements d'Avenir} program (ANR-17-EURE-0010) is gratefully acknowledged.}}
\author{Vasilii Ivanik\thanks{ICEF, NRU HSE, 11 Pokrovsky Boulevard, 109028 Moscow, Russia. Email: \href{mailto:vsivanik@edu.hse.ru}{\texttt{vsivanik@edu.hse.ru}}.}
\and
Georgy Lukyanov\thanks{Corresponding author. Toulouse School of Economics, 1, Esplanade de l'Universit\'e, 31080 Toulouse Cedex 06, France. Email: \href{mailto:georgy.lukyanov@tse-fr.eu}{\texttt{georgy.lukyanov@tse-fr.eu}}.}}
\date{14 September 2026}

\begin{document}
\maketitle

\begin{abstract}
We study social learning when agents choose costly information and prefer less popular actions. Popularity changes decision cutoffs and can restore the value of information. Under diminishing, nonsummable popularity updates, we characterize complete belief learning by positive attainable net information value at every interior belief, allowing vanishing entry fees and experiments approaching no information. With Gaussian signals and power precision costs, learning is complete exactly when contrarian incentives can align every belief and costs vanish faster than the square root of precision. Under weaker incentives, optimal purchases can have a positive precision floor and a finite expected count even without an entry fee. Rapid popularity updating can also stop learning despite profitable alignment. A positive fixed fee uniformly bounds purchases; a binary illustration characterizes terminal errors and separates belief accuracy from the frequency of correct actions.
\end{abstract}



\section{Introduction}\label{sec:introduction}

An analyst who is about to recommend an investment can draw on the
recommendations made by others. If the existing consensus is sufficiently
informative, there may be little reason to undertake an additional
investigation. At the same time, the analyst may benefit from offering a
recommendation that differs from the consensus. Such an incentive could
encourage further research, but it could also induce disagreement without
producing any new information. We ask under what conditions a preference for
less popular actions sustains the acquisition of information in a society
where agents learn from one another.

To study this question, we consider a sequence of agents who choose how much
private information to acquire before taking a binary action. Each agent
values making the correct decision and receives a premium for choosing an
action that was less popular among her predecessors. In the analyst
interpretation, the two actions correspond to favorable and unfavorable
recommendations, while information acquisition represents the research
undertaken before issuing a recommendation. The premium is specified directly
in the agent's payoff; deriving it from a market for advice is outside the
scope of the model.\footnote{Forecasting and advice
provide motivations for such a premium. \citet{LasterEtAl1999} and
\citet{OttavianiSorensen2006} study incentives to differentiate forecasts,
while \citet{BernhardtEtAl2006} document anti-herding in analysts' earnings
forecasts. Here the decision is binary, and an analyst's reward for a
distinctive recommendation is specified directly.}
An equivalent interpretation is a congestion cost. The proportional minority
premium differs from the linear congestion payoff of \citet{EysterEtAl2014}
only by an action-independent constant.

All agents observe the full history of actions, which affects their decisions
through two channels. First, earlier actions convey information about the
state and thus determine the public belief. Second, the relative popularity
of the two actions affects their payoffs. These channels need to be kept
separate. When an agent acquires no information, her action is predictable
from the public history and does not change the belief. It still enters the
popularity index, however, and can therefore change the next agent's incentive
to acquire information.

The mechanism works through the agent's indifference between the two actions.
Information is most valuable when, in its absence, she is indifferent: a
private signal can then change her decision in either direction. During a
spell without acquisition, the changes in popularity bring the decision
cutoff closer to the public belief, as far as the strength of the premium
allows. Further investigation may consequently become worthwhile even though
the public has learned nothing during the intervening periods. A stronger
premium expands the range of cutoffs that popularity can induce. Its effect
at a particular history is more qualified, since moving the cutoff beyond
indifference reduces the value of information.

Our main result gives a necessary and sufficient condition for this mechanism
to deliver complete belief learning from every interior prior and initial
popularity index. At each interior belief, some experiment must have positive
net value at the most favorable cutoff that popularity can induce. We allow
informative experiments to approach the choice of no information, and any
additional nonnegative entry fee may converge to zero at an arbitrary rate.
The popularity index is updated in diminishing steps whose sum is infinite.
Within this setting, the condition is stated entirely in terms of the
available experiments and their costs. The informativeness of the experiments
actually purchased is determined in equilibrium.

Establishing the result requires allowing for agents who continue to acquire
information but choose progressively weaker experiments. In this case, there
need not be a final pause during which the belief is fixed. We show that
popularity nevertheless converges to the level most favorable for information
acquisition at the limiting belief. If some experiment remained strictly
profitable there, optimal choices would continue to convey a positive amount
of information through actions. Such a limiting belief is impossible: the
total expected reduction in public uncertainty is bounded by the uncertainty
in the prior. This argument also gives a general bound relating the private
value of information to the movement in public beliefs.

We then apply the criterion to Gaussian signals, with information cost given
by a power of precision. This specification allows us to distinguish the role
of the cost function from that of contrarian incentives. At indifference, the
value of a weak signal is proportional to the square root of precision.
Complete learning requires costs to vanish faster than this gain, together
with a premium strong enough to bring the agent to indifference at every
public belief. In particular, a linear precision cost permits complete
learning despite having a positive marginal cost at zero. The reason is that
a small increase in precision has a relatively large effect on the value of
a binary decision when the agent is initially indifferent.

If the premium is too small, the agent cannot be brought to indifference when
the public belief is close to certainty. A weak signal would change her
decision only after a very unlikely realization, and the resulting gain fails
to cover its cost. Beliefs sufficiently close to certainty are then absorbing.
Moreover, in the case where weak experiments are profitable at indifference,
we show that weak congestion imposes a positive lower bound on the precision
of optimally purchased signals. The expected number of purchases is therefore
finite even without an entry fee. Although arbitrarily cheap experiments are
available, choosing arbitrarily small purchases is not optimal.

The way popularity is updated also matters. We construct an example with
linear precision cost, no entry fee, and a symmetric initial belief and
popularity index. When popularity is measured by the empirical frequency of
past actions, learning is complete. When each new action receives a large,
constant weight, the first recommendation changes popularity so much that no
subsequent agent finds information worth purchasing. Actions then alternate
forever without changing the public belief. The comparison holds the
observation structure fixed: in both cases, agents have access to the entire
history. Giving recent recommendations greater weight in payoffs can thus
stop the production of information even when the earlier record remains
available.

Finally, we consider an entry fee that remains strictly positive. For any
experiment menu and public cutoff, the expected number of purchases is bounded
by prior variance divided by the squared fee. Learning is then incomplete
provided informative actions have full support. With a fixed binary signal,
we obtain explicit expressions for the beliefs at which acquisition restarts,
the terminal error, and the number of purchases. This illustration also shows
why more accurate beliefs need not imply more accurate actions. Stronger
contrarian incentives initially improve both measures, but eventually induce
agents to act against their terminal belief without acquiring further
information. We compare these objective measures of accuracy throughout;
a welfare comparison across different preferences would require an additional
criterion.

The paper builds on the proportional-congestion model of
\citet{EysterEtAl2014}. With exogenously supplied signals, they establish the
possibility of temporary cascades, the threshold for complete learning, and
the decline in the frequency of correct actions under strong congestion. They
also show that uninformative cycling can prevent learning when congestion
changes fail to vanish. We
retain their payoff structure and examine the incentive to produce the
information from which others learn. This distinction matters particularly
when agents can choose experiments arbitrarily close to no information. The
argument concerning popularity is related to their analysis of vanishing
changes in congestion. Endogenous experiment choice additionally determines
the cost boundary for learning and, under weaker congestion, the lower bound
on purchased precision.

Costly information acquisition in social learning is studied by
\citet{BurguetVives2000}, \citet{Yang2011}, \citet{MuellerFrankPai2016}, and
\citet{Ali2018}. Ali treats the null choice separately from a compact menu of
informative experiments. Our continuous-precision menu includes informative
experiments converging to the null: every fixed positive precision costs
something, yet no common positive cost floor exists. In the smooth prediction
environment of Burguet and Vives, positive marginal information cost at zero
prevents unbounded accumulation. Here the square-root gain at a binary
decision cutoff explains why linear precision cost can sustain learning.
This comparison concerns the incentive to purchase weak information; the two
models also differ in what agents observe about earlier decisions.\footnote{Other information
choices give different margins. \citet{HendricksEtAl2012} study costly search
with aggregate purchase histories; \citet{BobkovaMass2022} study allocation of
an information budget between common and idiosyncratic payoff components.}

Our characterization uses the stationary-belief logic developed in the
social-learning literature. In particular, Theorem~1 of
\citet{KartikEtAl2024} relates adequate learning from every prior to adequate
knowledge at every stationary belief, under fixed preferences and exogenous
information. Here the additional step is to establish where the payoff cutoff
converges while agents choose their experiments, including experiments that
approach the null. The resulting condition concerns net information value at
that attainable cutoff. Related work considers collective
preferences and payoff externalities \citep{AliKartik2012,Arieli2017}, a desire
to differ \citep{Heinsalu2019}, competition for followers
\citep{Song2025}, and socially optimal contrarian experimentation
\citep{SmithSorensenTian2021}. The present payoff premium is private and
specified directly.\footnote{Throughout the paper, recent actions may receive
more weight in payoffs while all past actions remain observable. This differs
from changing the observation network, as in \citet{AcemogluEtAl2011}, or the
selection of consumers who generate reviews, as in
\citet{BesbesScarsini2018}. \citet{BikhchandaniEtAl2024} survey the broader
observational-learning literature.}

Section~\ref{sec:model} introduces the model and information-value argument.
Section~\ref{sec:continued} establishes the learning criterion.
Section~\ref{sec:gaussian} derives the Gaussian boundary and the fast-updating
example. Section~\ref{sec:fixed} studies fixed fees and the binary illustration.
Section~\ref{sec:conclusion} concludes. Proofs and supplementary binary
calculations follow in the appendices.

\section{Model and the value of information}\label{sec:model}

\subsection{Actions, beliefs, and popularity}

There is a fixed state $\theta\in\{0,1\}$ and a common prior
$\mu_1\in(0,1)$. A sequence of agents, indexed by $t=1,2,\ldots$, each
make one decision. Before choosing her experiment, agent $t$ observes all
earlier actions and forms the public belief
\begin{equation}\label{eq:public-belief}
 \mu_t=\Prb(\theta=1\mid\mathcal H_{t-1}),\qquad
 \mathcal H_{t-1}=\sigma(a_1,\ldots,a_{t-1}).
\end{equation}
Popularity is a public index $z_t\in[0,1]$ satisfying
\begin{equation}\label{eq:gain-update}
 z_{t+1}=z_t+\gamma_t(a_t-z_t),\qquad 0<\gamma_t\leq1,
\end{equation}
with a deterministic gain sequence. Except in the constant-gain example, we
assume
\begin{equation}\label{eq:gain-conditions}
 \gamma_t\to0,\qquad\sum_{t\geq1}\gamma_t=\infty.
\end{equation}
The empirical-frequency case, denoted by $p_t$, uses $\gamma_t=1/t$:
\begin{equation}\label{eq:popularity}
 p_1=\tfrac12,\qquad p_t=\frac1{t-1}\sum_{i<t}a_i\quad(t\geq2).
\end{equation}
Allowing other diminishing gains accommodates a greater weight on recent
actions in the popularity index. Throughout, the agent can still consult all
past actions when forming her belief.

After observing history, the agent chooses an experiment and observes its
private realization. She then takes a public action $a_t\in\{0,1\}$. Write
$z_t(1)=z_t$ and $z_t(0)=1-z_t$. Her payoff before information costs is
\begin{equation}\label{eq:congestion-payoff}
 \1\{a_t=\theta\}-kz_t(a_t),\qquad k\geq0.
\end{equation}
Equivalently, adding the action-independent constant $k$ gives a minority
premium $k(1-z_t(a_t))$. Popularity depends only on actions already taken,
so the agent treats the index as given when making her decision. Since she
acts once, she also has no incentive to influence the decisions of her
successors.

At private posterior $q$, the difference between the gross payoffs from actions
1 and 0 is $2(q-c_k(z))$, where
\begin{equation}\label{eq:cutoff}
 c_k(z)=\frac12+k\left(z-\frac12\right).
\end{equation}
The optimal action is therefore $a=\1\{q\geq c_k(z)\}$. Maximized gross
payoff is
\begin{equation}\label{eq:max-payoff}
 \max\{q-kz,1-q-k(1-z)\}=\frac{1-k}{2}+|q-c_k(z)|.
\end{equation}
A signal can change the decision only if the cutoff is interior. Notice also
that the public belief and the popularity index generally differ. The belief
uses the likelihood of the observed history, whereas popularity records a
weighted frequency of actions. Even a history that reveals little about the
state can therefore have a substantial effect on the agent's payoff.

\subsection{Experiments and equilibrium}

At public belief $\mu$, experiment $e$ generates a private posterior $Q$
with law $\nu_{\mu,e}$ and costs $C(e)$. In addition to this base cost, a
non-null experiment incurs a publicly known, deterministic entry fee
$F_t\geq0$. The null experiment $e_0$ leaves the posterior at $Q=\mu$
and has zero total cost. The agent chooses her experiment before receiving
any private information, and there are no privately observed cost types.

\begin{assumption}\label{ass:continuous-menu}
The experiment set $\mathcal E$ is a fixed compact metric space containing
$e_0$. The laws $\nu_{\mu,e}$ are jointly weakly continuous in $(\mu,e)$,
have mean $\mu$, and satisfy $\nu_{\mu,e_0}=\delta_\mu$. Base cost $C$ is
continuous and nonnegative, with $C(e_0)=0$. The technology, costs, and
selection rule are public. The menu and base costs do not depend on the cutoff.
\end{assumption}

Bayes plausibility is conditional at each history:
$\E[Q_t\mid\mathcal H_{t-1}]=\mu_t$. It is not a restriction only on the
unconditional signal distribution.\footnote{For $\mu\in(0,1)$, a mean-$\mu$
posterior law is implementable using state-contingent laws
$\nu_{\mu,e}(dq\mid\theta=1)=(q/\mu)\nu_{\mu,e}(dq)$ and
$\nu_{\mu,e}(dq\mid\theta=0)=((1-q)/(1-\mu))\nu_{\mu,e}(dq)$.
These formulas specify how private information is related to the state.
The general formulation permits this implementation to depend on the public
belief. The Gaussian and binary specifications below use signal laws that,
conditional on the state, are independent of the prior.}
The compact menu may contain informative experiments converging to the null.
No positive lower bound on their cost or informativeness is imposed.

\begin{assumption}\label{ass:ties}
Action ties select action 1. Acquisition ties select the null experiment.
Ties among non-null maximizers are resolved by a fixed deterministic rule.
\end{assumption}

A pure-strategy perfect Bayesian equilibrium requires optimal experiment and
action choices at every history, together with Bayes updating wherever
applicable. The acting agent solves a static problem. Observers infer her
experiment from the public history and selection rule, so the experiment need
not be separately observed. A null action is predictable and leaves the public
belief unchanged. A purchased experiment need not be fully revealed by the
action.

After a zero-probability public action, assign the next agent a public belief
and apply the same optimal rules from that history onward. An agent who
deviates in her experiment choice updates from her pre-experiment belief using
her actual experiment. The public belief assigned after her action does not
replace this private posterior and cannot reward or deter the deviation,
because she has no continuation payoff. The maximization in
\eqref{eq:p3-W} is attained by compactness and continuity. If its value exceeds
the entry fee, a maximizing experiment is non-null and optimal to purchase;
otherwise the null is optimal. Thus the fee discontinuity at the null does
not prevent attainment. These rules recursively define an equilibrium and,
given the selection convention, its on-path outcome. Our conclusions concern
that outcome.

Complete belief learning means $\mu_t\to\theta$ almost surely. We examine
the limiting frequency of correct actions separately: an agent with accurate
beliefs may still prefer the less popular, incorrect action.

\subsection{Information value and what actions reveal}

By \eqref{eq:max-payoff}, gross information value and optimized base net value
are
\begin{align}
 D_e(\mu,c)&=\E_{\nu_{\mu,e}}|Q-c|-|\mu-c|,\label{eq:p3-D}\\
 W(\mu,c)&=\max_{e\in\mathcal E}\{D_e(\mu,c)-C(e)\}.\label{eq:p3-W}
\end{align}
The agent purchases information exactly when $W(\mu_t,c_t)>F_t$.
The null experiment makes $W$ nonnegative.

\begin{lemma}\label{lem:general-value}
For every experiment,
\begin{equation}\label{eq:general-D-options}
 D_e(\mu,c)=
 \begin{cases}
  2\E(c-Q)_+,&c\leq\mu,\\
  2\E(Q-c)_+,&c\geq\mu.
 \end{cases}
\end{equation}
Both $D_e$ and $W$ are weakly increasing in $c$ up to $\mu$ and weakly
decreasing thereafter. Moreover,
\begin{equation}\label{eq:universal-bound}
 0\leq D_e(\mu,c)\leq2\mu(1-\mu),
\end{equation}
and $W$ is continuous under Assumption~\ref{ass:continuous-menu}.
\end{lemma}

\begin{proof}
The identities follow by splitting the absolute value at $c$ and using
$\E Q=\mu$. They give the common mode at $c=\mu$, which is preserved by
subtracting a cutoff-independent cost and maximizing. At the mode, convexity
gives $|Q-\mu|\leq Q(1-\mu)+(1-Q)\mu$; taking expectations proves the
bound. Joint weak continuity and continuity of $C$ make the objective jointly
continuous on the compact menu. The maximum theorem gives continuity of $W$.
\end{proof}

To understand the lemma, consider an agent who would choose action 1 using
public information alone. The signal creates value only through realizations
that lead her to choose action 0 instead. As the cutoff approaches the public
belief, such realizations become more valuable; once the cutoff passes the
belief, the relevant gain comes from switching in the opposite direction.
Information value is therefore greatest at indifference for every experiment.
A stronger contrarian incentive raises information demand at a given history
only insofar as it moves the cutoff toward this point.

The same threshold structure links private information value to public
learning. Let $D_t$ denote the conditional gross value of the experiment
chosen at date $t$, with $D_t=0$ for the null.

\begin{lemma}\label{lem:value-budget}
For any public cutoff rule and deterministic experiment choice,
\begin{equation}\label{eq:p3-budget}
 \E\sum_{t\geq1}D_t^2
 \leq\mu_1(1-\mu_1)-\E[\mu_\infty(1-\mu_\infty)]
 \leq\mu_1(1-\mu_1).
\end{equation}
In particular, $D_t\to0$ almost surely. This bound does not require
vanishing popularity gains, a stationary menu, or a positive entry fee.
\end{lemma}

\begin{proof}
Put $\Delta\mu_t=\mu_{t+1}-\mu_t$. Conditional on public history, the
cutoff is known and $a_t=\1\{Q_t\geq c_t\}$. Bayesian updating gives
$\mu_{t+1}=\E[Q_t\mid\mathcal H_t]$. On each action event, $Q_t-c_t$
has a constant weak sign. Consequently,
\[
 \E[|Q_t-c_t|\mid\mathcal H_{t-1}]
 =\E[|\mu_{t+1}-c_t|\mid\mathcal H_{t-1}],
\]
and the triangle and Cauchy--Schwarz inequalities imply
\begin{equation}\label{eq:p3-budget-local}
 D_t\leq\E[|\Delta\mu_t|\mid\mathcal H_{t-1}]
 \leq\left(\E[(\Delta\mu_t)^2\mid\mathcal H_{t-1}]\right)^{1/2}.
\end{equation}
The bounded posterior martingale converges in $L^2$. Orthogonality of its
increments gives
\[
 \E\sum_t(\Delta\mu_t)^2
 =\E\mu_\infty^2-\mu_1^2
 =\mu_1(1-\mu_1)-\E[\mu_\infty(1-\mu_\infty)].
\]
Summing \eqref{eq:p3-budget-local} proves the result. Finite expected sum
implies $\sum_t D_t^2<\infty$ almost surely.
\end{proof}

The link between private value and public learning does not require the
action to reveal the private posterior. Observers may be unable to distinguish
several signal realizations that induce the same action. All these
realizations nevertheless lie on the same linear branch of the payoff, so
averaging over them preserves gross information value. Thus, an experiment
that is sufficiently valuable to the acting agent must also produce a
corresponding movement in public beliefs. The bounded amount of prior
uncertainty limits how often this can happen.

\section{When learning continues}\label{sec:continued}

\subsection{Popularity while information is still acquired}

As popularity varies over $[0,1]$, the feasible cutoff interval is
$C_k=[(1-k)/2,(1+k)/2]$. For $k>0$, define
\begin{align}
 p^*(\mu,k)&=\proj_{[0,1]}\left(\frac12+\frac{\mu-1/2}{k}\right),
 \label{eq:pstar}\\
 \widehat c_k(\mu)&=\proj_{C_k}\mu.\label{eq:chat}
\end{align}
Here $\proj_{[L,U]}x=\min\{U,\max\{L,x\}\}$. For $k=0$,
$\widehat c_0(\mu)=1/2$.

\begin{lemma}\label{lem:ongoing-tracking}
The public posterior converges almost surely to a limit $\mu_\infty$. For
$k>0$, even if information is acquired infinitely often,
\[
 z_t\longrightarrow p^*(\mu_\infty,k),\qquad
 c_t\longrightarrow\widehat c_k(\mu_\infty)
 \quad\text{almost surely}.
\]
The cutoff conclusion also holds for $k=0$.
\end{lemma}

The lemma applies even if there is no final purchase. The reason is that the
action itself tells observers on which side of the cutoff the private
posterior lies. Action 1 implies $\mu_{t+1}\geq c_t$, while action 0
implies $\mu_{t+1}<c_t$. Hence, for $k>0$, the realized path satisfies
\[
 a_t=\1\left\{z_t\leq\frac12+\frac{\mu_{t+1}-1/2}{k}\right\}.
\]
This identity describes the outcome after the action is observed; the acting
agent does not know the subsequent public posterior in advance. As public
beliefs converge, the threshold on the right also converges. Each action
moves popularity toward that threshold, and the diminishing size of the
updates prevents persistent overshooting. The argument, including the case
of an endpoint target, is given in Appendix~\ref{app:continued-proofs}. When $k=0$,
popularity has no effect on payoffs: after acquisition stops at belief $\mu$,
all actions equal $\1\{\mu\geq1/2\}$.

\subsection{A criterion in terms of attainable information value}

Define
\begin{equation}\label{eq:p3-G}
 G_k(\mu)=W(\mu,\widehat c_k(\mu)),\qquad
 Z_k=\{\mu\in[0,1]:G_k(\mu)=0\}.
\end{equation}
Lemma~\ref{lem:general-value} implies
$G_k(\mu)=\max_{z\in[0,1]}W(\mu,c_k(z))$. It is the best net experiment
value that popularity can induce, before the entry fee.

\begin{theorem}\label{thm:continued-learning}
Under Assumption~\ref{ass:continuous-menu}, \eqref{eq:gain-conditions}, and $F_t\to0$,
\[
 \Prb(\mu_\infty\in Z_k)=1.
\]
For the given fee and gain sequences, complete belief learning occurs from
\emph{every} interior prior and \emph{every} initial popularity index if and
only if
\begin{equation}\label{eq:p3-criterion}
 G_k(\mu)>0\qquad\text{for every }\mu\in(0,1).
\end{equation}
No rate restriction on $F_t\to0$ is required. For $k\geq1$, the criterion
reduces to $W(\mu,\mu)>0$ at every interior belief.
\end{theorem}

\begin{proof}
On a purchase date, the chosen experiment maximizes $D_e-C(e)$, so
$W(\mu_t,c_t)=D_t-C(e_t)\leq D_t$. On a null date, optimality gives
$W(\mu_t,c_t)\leq F_t$. Therefore at every date,
\[
 0\leq W(\mu_t,c_t)\leq D_t+F_t.
\]
Equation~\eqref{eq:p3-budget} and $F_t\to0$ imply
$W(\mu_t,c_t)\to0$ almost surely. Lemma~\ref{lem:ongoing-tracking} and
continuity then imply $G_k(\mu_\infty)=0$.

If $G_k$ is positive at every interior belief, its zero set is contained in
$\{0,1\}$. Since $\mu_\infty=\E[\theta\mid\mathcal H_\infty]$,
\[
 \E[(\theta-\mu_\infty)^2]=\E[\mu_\infty(1-\mu_\infty)]=0,
\]
and $\mu_\infty=\theta$ almost surely. This applies to every interior prior
and initial index. Conversely, if $G_k(\mu_0)=0$ for some interior $\mu_0$,
then $W(\mu_0,c_k(z))=0$ for every $z$. No non-null experiment is strictly
profitable at any nonnegative fee. Starting at $\mu_0$, the tie convention
therefore selects null forever, so learning fails. Finally, $C_k$ contains
$[0,1]$ when $k\geq1$, making its projection of $\mu$ equal to $\mu$.
\end{proof}

The necessity part concerns learning from every interior prior. If $G_k$
vanishes at an interior belief, an economy starting at that belief never
acquires information. Another prior may still lead to complete learning if
this obstruction cannot be reached. For sufficiency, no lower bound on the
informativeness of chosen experiments is required. Their gross values tend
to zero by Lemma~\ref{lem:value-budget}, while popularity brings the cutoff
to its most favorable attainable position. A limiting interior belief at
which information remains profitable would be inconsistent with these two
properties.

When $k\geq1$, popularity can induce indifference at every belief. The
condition then reduces to $W(\mu,\mu)>0$ throughout the interior. For
smaller $k$, the range of feasible cutoffs is itself restrictive: near
certainty, no popularity index brings the agent to indifference. The Gaussian
specification below shows how this restriction interacts with the cost of
weak information.

\section{Gaussian precision and the cost of weak information}\label{sec:gaussian}

\subsection{The learning boundary}

An experiment of precision $\rho>0$ generates
\[
 s\mid\theta\sim N(\theta,1/\rho),
\]
and has base cost $C(\rho)=\kappa\rho^\beta$, where $\kappa>0$ and
$\beta>0$. Precision zero denotes no information. Throughout this section,
the additional entry fee is nonnegative and converges to zero; it may be
identically zero. At an aligned cutoff, the signal threshold is $s=1/2$ and
\begin{equation}\label{eq:gaussian-value}
 D_G(\mu,\mu;\rho)
 =2\mu(1-\mu)\left[2\PhiN(\sqrt\rho/2)-1\right],
\end{equation}
where $\PhiN$ and $\phi$ are the standard normal distribution function and
density. Since gross value is at most $1/2$, no precision with cost above
$1/2$ is optimal. We can therefore restrict the menu to a compact interval
including zero. The posterior laws satisfy
Assumption~\ref{ass:continuous-menu}.

\Needspace{12\baselineskip}
\begin{theorem}\label{thm:gaussian-power}
Under full-history observation and vanishing, nonsummable popularity gains,
complete belief learning with Gaussian signals and power precision costs
occurs from every interior prior and every initial index exactly when
\begin{equation}\label{eq:power-boundary}
 k\geq1\quad\text{and}\quad\beta>\frac12.
\end{equation}
More precisely:
\begin{enumerate}[label=(\roman*)]
 \item If \eqref{eq:power-boundary} holds, $\mu_t\to\theta$ and there are
 infinitely many purchases almost surely.
 \item If $k<1$ or $\beta\leq1/2$, then $\mu_\infty\in(0,1)$ almost surely
 from every interior prior.
 \item If $k<1$ and $\beta>1/2$, there is a constant
 $\underline\rho(k,\kappa,\beta)>0$ such that every optimally purchased
 experiment has $\rho\geq\underline\rho$. Consequently,
 \[
 \E N_I\leq
 \frac{\mu_1(1-\mu_1)}{\kappa^2\underline\rho^{\,2\beta}}<\infty.
 \]
 This lower bound on purchased precision is endogenous; the feasible menu
 includes arbitrarily small positive precisions.
\end{enumerate}
\end{theorem}

Figure~\ref{fig:cost-regions} summarizes the learning and purchase outcomes.

\begin{figure}[!htbp]
\centering
\begin{tikzpicture}[x=4.5cm,y=3.5cm,>=Latex]
 \definecolor{learnblue}{RGB}{35,100,122}
 \definecolor{stoporange}{RGB}{161,102,44}
 \fill[stoporange!12] (0,.5) rectangle (1,1.5);
 \fill[learnblue!12] (1,.5) rectangle (2.12,1.5);
 \fill[black!7] (0,0) rectangle (2.12,.5);
 \draw[black!55,line width=.8pt] (0,.5)--(2.12,.5);
 \draw[learnblue,line width=1.1pt] (1,.5)--(1,1.5);
 \draw[black!70,->,line width=.7pt] (0,0)--(2.22,0);
 \draw[black!70,->,line width=.7pt] (0,0)--(0,1.59);
 \foreach \x/\lab in {1/1,2/2} {
   \draw[black!70] (\x,0)--(\x,-.025);
   \node[below,font=\small] at (\x,-.025) {$\lab$};}
 \foreach \y/\lab in {.5/{\tfrac12},1/1} {
   \draw[black!70] (0,\y)--(-.025,\y);
   \node[left,font=\small] at (-.025,\y) {$\lab$};}
 \node[below left,font=\small] at (0,0) {$0$};
 \node[below,font=\small] at (1.1,-.14) {contrarian intensity $k$};
 \node[rotate=90,font=\small] at (-.24,.82) {precision-cost exponent $\beta$};
 \node[align=center,text width=4.0cm,font=\small,text=stoporange!80!black]
   at (.50,1.15) {\textbf{Incomplete beliefs}\\[4pt]
   Finite expected purchases\\[2pt]Positive precision floor};
 \node[align=center,text width=4.4cm,font=\small,text=learnblue!85!black]
   at (1.59,1.15) {\textbf{Complete beliefs}\\[4pt]
   Infinitely many purchases};
 \node[align=center,font=\small,text=black!80] at (1.06,.25)
   {\textbf{Incomplete beliefs}\\[4pt]Purchase count not classified};
 \fill[learnblue] (1,1) circle[radius=1.6pt];
 \node[anchor=west,font=\small,text=learnblue!85!black] at (1.035,.93) {$P$};
\end{tikzpicture}
\caption{Gaussian learning with zero or vanishing entry fees and diminishing, nonsummable
popularity gains. The line $\beta=1/2$ belongs to the lower region; $k=1$
belongs to the complete-learning region above it. At $P$, constant-gain
updating instead stops acquisition after one purchase
(Proposition~\ref{prop:gaussian-cycle}).}
\label{fig:cost-regions}
\end{figure}
\FloatBarrier

Figure~\ref{fig:cost-regions} separates the restrictions on cost and congestion.
The two incomplete-learning regions have different purchase implications.
When weak information is worthwhile at alignment but congestion is too weak,
optimal precision stays away from zero. Below the cost boundary, the theorem
establishes incomplete learning without a general conclusion about the number
of purchases.

At every interior aligned belief,
\begin{equation}\label{eq:gaussian-small}
 D_G(\mu,\mu;\rho)
 =2\phi(0)\mu(1-\mu)\sqrt\rho+O(\rho^{3/2}).
\end{equation}
For $\beta>1/2$, the cost of a sufficiently small positive precision is
lower than its value at indifference. If congestion allows this alignment at
every belief, Theorem~\ref{thm:continued-learning} therefore yields complete
learning. Linear precision cost satisfies this condition, even though its
marginal cost at zero is positive. This observation depends on measuring the
experiment by its precision. In signal-strength units $\eta=\sqrt\rho$,
the same comparison is between cost $\kappa\eta^{2\beta}$ and a gain of
order $\eta$. In either parametrization, weak information is worth purchasing
because its cost vanishes faster than its value.

There are two reasons why learning can fail. If $\beta\leq1/2$, information
is too costly near certainty even when the agent is indifferent. If $k<1$,
a sufficiently extreme belief lies outside the feasible cutoff interval.
Starting from such a belief, a very weak signal changes the decision only
after a realization in a Gaussian tail, whose contribution cannot cover a
power cost. Each restriction creates absorbing neighborhoods of certainty.
Because Gaussian signals have full support, every posterior reached in finite
time remains interior; the absorbing neighborhoods then prevent convergence
to certainty.

The lower bound on purchased precision when $k<1$ and $\beta>1/2$ is
implied by optimal choice. Since purchases take place away from the absorbing
tails, a sequence of purchased precisions tending to zero would have to
approach either an aligned or a non-aligned cutoff. In the first case, another
experiment would remain strictly profitable and would be preferred to the
vanishing purchase. In the second, the proposed weak experiment would not
cover its cost. Neither case is consistent with optimality. Every purchase
therefore uses a positive amount of the public information budget, which
bounds the expected number of purchases. The details are in
Appendix~\ref{app:continued-proofs}.

\subsection{Beliefs and actions under complete learning}

For empirical popularity, Lemma~\ref{lem:ongoing-tracking} also gives a direct
behavioral implication. If beliefs converge to the true state and $k\geq1$,
the limiting frequency of correct actions is, almost surely,
\begin{equation}\label{eq:p3-complete-accuracy}
 \frac{1+k}{2k}.
\end{equation}
Indeed, the limiting action-1 frequency is
$1/2+(\theta-1/2)/k$. Correct actions have limiting frequency one at $k=1$,
but a smaller frequency for every $k>1$. Complete public beliefs therefore do
not eliminate disagreement generated by the payoff premium. This distinction
is already present in the exogenous-information benchmark of
\citet{EysterEtAl2014}; it remains here when information is chosen and paid for.

\subsection{When popularity adjusts too quickly}

We now examine why the size of popularity updates matters for learning.
Consider the point $P$ in Figure~\ref{fig:cost-regions}, with linear
precision cost and the weakest congestion strength that permits alignment
at every belief. These primitives satisfy the learning condition under
diminishing, nonsummable gains. The next example shows what changes when
recent actions instead receive a large, constant weight.

\begin{proposition}\label{prop:gaussian-cycle}
Let $k=1$, $C(\rho)=\rho$, $F_t=0$, and use constant gain $\gamma_t=4/5$.
Start at $\mu_1=z_1=1/2$. The first agent buys the unique aligned precision
$\rho_*>0$ satisfying
\[
 \sqrt{\rho_*}=\frac14\phi(\sqrt{\rho_*}/2).
\]
After her action, the public belief is $r_*$ or $1-r_*$, where
$r_*=\Phi_N(\sqrt{\rho_*}/2)\in(1/2,0.52)$. No further agent acquires
information. Actions alternate, and popularity approaches the two-cycle
$\{1/6,5/6\}$. Beliefs remain interior and the long-run correct-action
frequency is $1/2$, although $G_1(\mu)>0$ for every interior belief.
\end{proposition}

The first action moves popularity to 0.9 or 0.1, whereas the public belief
moves only to approximately 0.51986 or its complement. The next agent
therefore faces a cutoff far from her belief and acquires no information.
Subsequent actions keep popularity alternating between two intervals in which
information cannot cover its cost. The proof verifies that these intervals
remain invariant, so the argument applies before the limiting cycle is
reached. If popularity were instead the empirical frequency, the same payoff
and information primitives would give complete learning and, by
\eqref{eq:p3-complete-accuracy}, a limiting correct-action frequency of one.

All past actions remain observable in this example. Rapid updating affects
learning through the incentive to acquire information, even though agents
retain access to everything their predecessors have revealed. Uninformative
cycling itself is already established by \citet{EysterEtAl2014}. The feature
here is that purchases stop despite the availability of Gaussian experiments
with unbounded likelihood ratios and no entry fee. In the analyst
interpretation, recent recommendations may attract greater commercial
attention while the full record remains available for research. Restricting
observation to a finite history would change inference as well as incentives;
that is a separate question not analyzed here.

\section{Fixed fees and a binary illustration}\label{sec:fixed}

\subsection{Finite acquisition and endogenous restarts}

Suppose now that every non-null experiment has an entry fee $F>0$ that does
not decline. The information budget immediately limits total acquisition.

\begin{theorem}\label{thm:finite-acquisition}
Suppose experiment choice is a deterministic function of public history,
every non-null experiment costs at least $F>0$, and actions follow the public
cutoff rule.  Let $N_I$ be the total number of non-null experiments.  For any
initial belief $\mu_1\in(0,1)$,
\begin{equation}\label{eq:finite-bound}
    \E[N_I]\leq\frac{\mu_1(1-\mu_1)}{F^2}.
\end{equation}
Only finitely many experiments are purchased almost surely.  This bound holds
for any cutoff determined by public history; it does not require a particular
popularity update or a stationary experiment menu.
\end{theorem}

\begin{proof}
On every purchase date, optimality gives $D_t>F$. If $b_t$ denotes purchase,
$F^2b_t\leq D_t^2$. Summing and applying Lemma~\ref{lem:value-budget}
proves \eqref{eq:finite-bound} and almost-sure finiteness.
\end{proof}

If every purchased experiment induces both actions with positive probability
under both states, finitely many purchases leave an interior posterior. Thus
complete belief learning fails for every $k$. Perfectly revealing experiments
are excluded from this conclusion, not from the count bound. In addition,
\eqref{eq:universal-bound} implies that no experiment covers the fee at a
belief satisfying $2\mu(1-\mu)\leq F$.

For the stationary menu of Assumption~\ref{ass:continuous-menu}, define
\begin{equation}\label{eq:Sk}
 S_k(F)=\{\mu\in[0,1]:G_k(\mu)\leq F\}.
\end{equation}

\begin{proposition}\label{thm:general-restart}
Under \eqref{eq:gain-conditions}, beliefs in $S_k(F)$ are absorbing. At a
belief outside $S_k(F)$, every no-acquisition spell ends in finite time.
The terminal belief belongs to $S_k(F)$ almost surely. The set contracts
weakly with $k$ and is independent of $k$ for $k\geq1$.
\end{proposition}

\begin{proof}
In $S_k(F)$, no feasible cutoff makes purchase worthwhile. Outside it, a
hypothetical infinite null spell would make the cutoff converge to
$\widehat c_k(\mu)$ by Lemma~\ref{lem:ongoing-tracking}. Continuity and
$G_k(\mu)>F$ would then imply profitable purchase, a contradiction.
Finitely many purchases imply that the limiting belief is in $S_k(F)$.
Finally, feasible cutoff intervals expand with $k$ and contain all beliefs
when $k\geq1$, proving the set comparisons.
\end{proof}

The stopping-set argument needs only attainment and continuity of
$W(\mu,\cdot)$ at each belief. It also applies to belief-dependent stationary
menus: each gross value is $2$-Lipschitz in the cutoff, and maximization
preserves that bound. Joint continuity in the belief is needed for the
continued-learning criterion, not for this fixed-belief restart argument.

Thus a pause in acquisition need not mean that learning has ended. Outside
the stopping set, changes in popularity eventually make another purchase
worthwhile. Inside it, even the most favorable attainable cutoff cannot
compensate for the fee. We next characterize this set and the resulting
terminal beliefs explicitly for a binary signal.

\subsection{A fixed binary signal}

Use empirical popularity, $\mu_1=p_1=1/2$, and a single binary experiment.
Conditional on the state, successive purchased signals are independent and
have accuracy
\begin{equation}\label{eq:binary-signal}
 \Prb(y_t=\theta\mid\theta)=r\in(1/2,1).
\end{equation}
There is no variable cost. Put $d=2r-1$.
\begin{assumption}\label{ass:binary}
The fixed fee satisfies $0<F<d/2$.
\end{assumption}
The signal is then worth purchasing at the symmetric initialization. Write
\begin{align}
 \pi_1(\mu)&=(1-r)+d\mu,&\pi_0(\mu)&=r-d\mu,
 \label{eq:signal-probs}\\
 q_1(\mu)&=\frac{\mu r}{\pi_1(\mu)},&
 q_0(\mu)&=\frac{\mu(1-r)}{\pi_0(\mu)}.
 \label{eq:binary-posteriors}
\end{align}
The posteriors satisfy $q_0<\mu<q_1$ at interior beliefs. Let $D_r$ be the
gross value of this signal. Its aligned value is $2d\mu(1-\mu)$, so the
intrinsic lower boundary is
\begin{equation}\label{eq:astar}
 a^*=\frac{1-\sqrt{1-2F/d}}2.
\end{equation}
No cutoff induces acquisition outside $(a^*,1-a^*)$. At lower congestion
strengths, the feasible cutoff interval can impose a tighter restriction.
Put $\ell_k=(1-k)/2$ and
$\overline D_k(\mu)=D_r(\mu,\widehat c_k(\mu))$.

\begin{proposition}\label{prop:binary-restart}
Under Assumption~\ref{ass:binary}, define
\begin{equation}\label{eq:ak}
a_k=
\begin{cases}
a^*,
    & \ell_k\leq a^*,\\[6pt]
\displaystyle
\frac{F/2+(1-r)\ell_k}{r-(2r-1)\ell_k},
    & \ell_k>a^*.
\end{cases}
\end{equation}
Then:
\begin{enumerate}[label=(\roman*)]
    \item $\overline D_k(\mu)>F$ if and only if
    $\mu\in(a_k,1-a_k)$.
    \item If $\mu\notin(a_k,1-a_k)$, purchasing the signal is unprofitable
    at every popularity history, so the belief is absorbing.
    \item If $\mu\in(a_k,1-a_k)$, any no-information spell is finite:
    popularity targeting eventually makes acquisition strictly profitable.
    \item The boundary $a_k$ is weakly decreasing in $k$ and reaches the
    intrinsic boundary $a^*$ once
    \begin{equation}\label{eq:k-saturation}
        k\geq 1-2a^*
        =\sqrt{1-\frac{2F}{2r-1}}.
    \end{equation}
    Stronger incentives have no further effect on the restart region beyond
    this point.
\end{enumerate}
\end{proposition}

At a belief in $(a_k,1-a_k)$, acquisition eventually restarts, although the
current popularity index may not yet make a purchase worthwhile. The wait can
be long when popularity is measured by empirical frequency: after a long
history, each additional action changes the index only slightly.
Appendix~\ref{app:binary-extra} derives the waiting-time formulas for sufficiently
long histories and explains the qualification needed for short ones.

Whenever the signal is purchased, positive value requires
$q_0(\mu)<c_t<q_1(\mu)$. The action therefore reveals the signal. Put
\begin{equation}\label{eq:lambda-x}
 \lambda=\log\frac r{1-r},\qquad x=\frac{1-r}{r}.
\end{equation}
At purchase dates, public log odds move by $+\lambda$ or $-\lambda$;
between them they remain fixed.

\begin{theorem}
\label{thm:binary-random-walk}
Let
\begin{equation}\label{eq:Hk}
    H_k=
    \left\lceil
    \frac{\log\bigl((1-a_k)/a_k\bigr)}{\lambda}
    \right\rceil.
\end{equation}
Starting from $\mu_1=1/2$, public log odds sampled at information dates form a
nearest-neighbor random walk on $\lambda\mathbb{Z}$, stopped upon first reaching
$-H_k\lambda$ or $H_k\lambda$.  Conditional on $\theta=1$, an upward step has
probability $r$.  Hence
\begin{align}
    \Prb(\text{wrong terminal belief}\mid\theta=1)
    &=\frac{x^{H_k}}{1+x^{H_k}}, \label{eq:error}\\
    \E[N_I\mid\theta=1]
    &=\frac{H_k(1-x^{H_k})}
    {(2r-1)(1+x^{H_k})}, \label{eq:expected-purchases}
\end{align}
where $N_I$ is the total number of signal purchases.  By symmetry, the same
formulas hold conditional on $\theta=0$.

Increasing $k$ weakly raises $H_k$, weakly lowers the terminal-belief error,
and weakly raises the expected number of purchased signals until the saturation
threshold in \eqref{eq:k-saturation} is reached.
\end{theorem}

Changing $k$ leaves the information supplied by each purchased signal
unchanged. Its effect operates through the stopping boundary, which determines
how many signals enter the public record before acquisition ends. Since this
boundary is reached on an integer lattice, terminal accuracy improves in
steps. Once the stopping boundary has been established, the expressions for
error and expected purchases follow from the standard random-walk calculation
given in Appendix~\ref{app:binary-proofs}.

\subsection{Belief accuracy and action accuracy}

Let
\begin{equation}\label{eq:uk}
 u_k=\frac1{1+x^{H_k}}
\end{equation}
be the upper terminal belief and the probability that the terminal belief
points toward the true state. At that upper belief, the limiting popularity
of action 1 is
\begin{equation}\label{eq:zk}
 z_k=\begin{cases}
 1,&k=0,\\
 \min\{1,\tfrac12+(u_k-1/2)/k\},&k>0.
 \end{cases}
\end{equation}

\begin{proposition}\label{prop:accuracy}
The limiting frequency of correct actions exists almost surely.  Its ex ante
expectation is
\begin{equation}\label{eq:Ak}
    A_k=u_kz_k+(1-u_k)(1-z_k).
\end{equation}
On any interval of $k$ over which $H_k$ is constant,
\begin{equation}\label{eq:Ak-piecewise}
    A_k=
    \begin{cases}
        u_k, & k\leq2u_k-1,\\[5pt]
        \displaystyle
        \frac12+\frac{(2u_k-1)^2}{2k},
        & k>2u_k-1.
    \end{cases}
\end{equation}
Therefore stronger contrarian incentives improve terminal-belief accuracy only
when they increase the stopping boundary $H_k$.  Once $H_k$ has saturated,
further increases in $k$ eventually reduce expected action accuracy toward
$1/2$ even though terminal-belief accuracy is unchanged.
\end{proposition}

Define
\begin{align}
 k_R&=1-2a^*,\label{eq:kR}\\
 H^*&=\left\lceil\frac{\log((1-a^*)/a^*)}{\lambda}\right\rceil,
 &u^*&=\frac1{1+x^{H^*}},&k_A&=2u^*-1.\label{eq:kA}
\end{align}

\begin{corollary}\label{cor:three-regimes}
The thresholds satisfy $0<k_R\leq k_A<1$.  Moreover:
\begin{enumerate}[label=(\roman*)]
    \item for $0\leq k<k_R$, the restart region expands with $k$,
    $H_k$ and expected information purchases weakly increase, terminal-belief
    error weakly decreases, and $A_k=u_k$ weakly increases;
    \item for $k_R\leq k\leq k_A$, the restart region, information purchases,
    terminal-belief error, and action accuracy are all constant, with
    $H_k=H^*$ and $A_k=u^*$;
    \item for $k>k_A$, information purchases and terminal beliefs remain
    constant, while $A_k=1/2+(2u^*-1)^2/(2k)$ decreases strictly toward
    $1/2$.
\end{enumerate}
For $k<k_A$, every action after information shutdown follows the terminal
belief's sign.  At $k=k_A$, the same is true apart from a possible one-period
tie-breaking exception at the lower terminal belief.  For $k>k_A$, some
agents deliberately choose against the terminal belief.
\end{corollary}

The two thresholds refer to different margins. At $k_R$, further increases
in the premium cease to expand the restart region. At $k_A$, the premium
becomes strong enough to induce actions against the terminal belief. Their
ordering allows an interval over which neither measure of accuracy changes,
although this interval can collapse to a point. The integer belief lattice
also matters: purchases and terminal accuracy can reach their final plateau
before the continuous restart region stops expanding at $k_R$.

For $r=2/3$ and $F=0.05$, the purchase boundary takes values 1 through 4.
The corresponding terminal errors are $1/3$, $1/5$, $1/9$, and $1/17$;
expected purchases are $1$, $3.6$, $7$, and $180/17$. The upward transitions
occur strictly above $9/80$, $11/24$, and $147/200$. Acquisition
indifference selects the lower value at each boundary.
Figure~\ref{fig:two-accuracies} shows the resulting steps and subsequent
divergence. Here $k_R\simeq0.83666$ and $k_A=15/17\simeq0.88235$.
Both accuracies reach $16/17$ before $k_R$; beyond $k_A$ only action accuracy
declines. These are objective-accuracy comparisons, not welfare claims.

\begin{figure}[htbp]
\centering
\begin{tikzpicture}
\begin{axis}[
 width=.94\textwidth,height=6.6cm,
 xmin=0,xmax=1.6,ymin=.60,ymax=.995,
 axis lines=left,axis line style={black!65},
 xlabel={contrarian intensity $k$},ylabel={long-run accuracy},
 label style={font=\small},tick label style={font=\small},
 tick align=outside,major tick length=2pt,
 xtick={0,.5,1,1.5},
 ytick={.6666667,.8,.8888889,.9411765},
 yticklabels={$\tfrac23$,$\tfrac45$,$\tfrac89$,$\tfrac{16}{17}$},
 legend style={at={(.985,.04)},anchor=south east,font=\small,
 draw=none,fill=white,inner sep=3pt},clip=false]
\fill[black!5] (axis cs:.836660027,.60) rectangle (axis cs:.882352941,.94117647);
\addplot[MidnightBlue!85!black,very thick] coordinates {(0,.6666667)(.1125,.6666667)};
\addlegendentry{belief accuracy $u_k$}
\addplot[Maroon!85!black,thick,densely dashed] coordinates {(0,.6666667)(.1125,.6666667)};
\addlegendentry{expected action accuracy $A_k$}
\addplot[MidnightBlue!85!black,very thick,forget plot] coordinates {(.1125,.8)(.458333333,.8)};
\addplot[MidnightBlue!85!black,very thick,forget plot] coordinates {(.458333333,.8888889)(.735,.8888889)};
\addplot[MidnightBlue!85!black,very thick,forget plot] coordinates {(.735,.94117647)(1.6,.94117647)};
\addplot[Maroon!85!black,thick,densely dashed,forget plot] coordinates {(.1125,.8)(.458333333,.8)};
\addplot[Maroon!85!black,thick,densely dashed,forget plot] coordinates {(.458333333,.8888889)(.735,.8888889)};
\addplot[Maroon!85!black,thick,densely dashed,forget plot] coordinates {(.735,.94117647)(.882352941,.94117647)};
\addplot[Maroon!85!black,thick,densely dashed,domain=.882352941:1.6,samples=90,forget plot]
 {.5+(15/17)^2/(2*x)};
\addplot[only marks,mark=*,mark size=2pt,MidnightBlue!85!black,forget plot]
 coordinates {(.1125,.6666667)(.458333333,.8)(.735,.8888889)};
\addplot[only marks,mark=*,mark size=2pt,mark options={fill=white},
 MidnightBlue!85!black,forget plot]
 coordinates {(.1125,.8)(.458333333,.8888889)(.735,.94117647)};
\draw[black!45,densely dotted] (axis cs:.836660027,.60)--(axis cs:.836660027,.97);
\draw[black!45,densely dotted] (axis cs:.882352941,.60)--(axis cs:.882352941,.97);
\node[anchor=south east,font=\small] at (axis cs:.827,.97) {$k_R$};
\node[anchor=south west,font=\small] at (axis cs:.892,.97) {$k_A$};
\end{axis}
\end{tikzpicture}
\caption{Belief accuracy and expected correct-action frequency ($r=2/3$,
$F=0.05$). Filled and open markers indicate included and excluded step
endpoints. Beliefs reach their plateau before $k_R$; beyond $k_A$, only
action accuracy declines.}
\label{fig:two-accuracies}
\end{figure}
\FloatBarrier

\subsection{The bridge to declining fees}

To relate the binary illustration to continued learning, keep the same signal
and let its fee follow a nonnegative deterministic sequence $F_t\to0$.
Here the cost declines along the history of a single economy. Early fees
may be arbitrarily high, and the decline need not be monotone.

\begin{proposition}\label{prop:declining-binary}
If $k\geq1$, complete belief learning occurs from every interior prior and
information is purchased infinitely often almost surely. If $0\leq k<1$,
define
\[
 \alpha_k=\frac{(1-r)(1-k)/2}{r-(2r-1)(1-k)/2}>0.
\]
Beliefs outside $(\alpha_k,1-\alpha_k)$ are absorbing. At a belief inside
that interval, any no-acquisition spell ends in finite calendar time.
Starting from $\mu_1=1/2$, the purchase-clock log odds are a random walk
stopped at $\pm H_k^0\lambda$, where
\[
 H_k^0=\left\lceil\frac{\log((1-\alpha_k)/\alpha_k)}\lambda\right\rceil.
\]
Conditional on either state,
\[
 \Prb(\text{wrong terminal belief}\mid\theta)
 =\frac{x^{H_k^0}}{1+x^{H_k^0}},\qquad
 \E[N_I\mid\theta]
 =\frac{H_k^0(1-x^{H_k^0})}{d(1+x^{H_k^0})}.
\]
Thus for $k<1$ purchases are finite and these terminal outcomes are
independent of the particular nonnegative fee sequence converging to zero.
The sequence can affect calendar delays.
\end{proposition}

For $k\geq1$, aligned value is positive at every interior belief, so the
general criterion applies. For $k<1$, the terminal law is that of the same
finite purchase-clock walk for every such fee sequence. The sequence changes
the waiting times between purchases. At $r=2/3$,
\[
 H_k^0=\left\lceil\log_2\frac{2(1+k)}{1-k}\right\rceil,
 \qquad k<1.
\]
Thus $H_k^0=2$ on $(0,1/3]$ and $H_k^0=3$ on $(1/3,3/5]$.
The endpoint convention is substantive: an optional experiment of zero gross
value is rejected even at zero fee.

A related exercise in Appendix~\ref{app:binary-extra} compares economies with
different stationary fees as the positive fee tends to zero. Both exercises
recover the familiar threshold for complete learning. In making the
comparison with free information, the acquisition convention must still be
respected: an optional experiment with zero value is declined, whereas an
exogenously supplied signal is received regardless of its value.

\section{Conclusion}\label{sec:conclusion}

We have studied how a preference for less popular actions affects the
information that agents choose to acquire. Such a preference changes the
value of investigation by moving the cutoff at which the agent switches
actions. Changes in popularity can make acquisition worthwhile again after
a pause, and their effect continues to be well defined when agents keep
purchasing progressively weaker signals.

The main result identifies when these incentives sustain complete belief
learning. Information must remain profitable at the most favorable attainable
cutoff at every interior belief. For Gaussian signals, this condition requires
both a sufficiently strong congestion effect and a cost that vanishes faster
than the value of weak information. Linear precision cost satisfies the latter
requirement. Under weaker congestion, however, optimal purchases can be bounded
away from zero in precision, giving a finite expected purchase count even
without an entry fee. Updating popularity too quickly provides a further
obstruction by preventing the cutoff from approaching the belief.

A strictly positive fixed fee always bounds the expected number of purchases.
The binary illustration shows how stronger contrarian incentives can then
improve terminal beliefs before they begin to reduce the frequency of correct
actions. This distinction limits the policy interpretation of the results.
Assessing a subsidy to minority positions would require a common welfare
criterion and an account of how the subsidy affects the underlying payoffs.
The analysis here concerns the information produced and the accuracy of
beliefs and actions under the specified incentives.

\appendix
\numberwithin{equation}{section}
\renewcommand{\thetheorem}{\Alph{section}.\arabic{theorem}}

\section{Proofs for continued learning and Gaussian precision}\label{app:continued-proofs}

\subsection{Proof of Lemma~\ref{lem:ongoing-tracking}}

\begin{proof}
Martingale convergence gives $\mu_t\to\mu_\infty$ almost surely. At every
positive-probability action event, conditional averaging of the private
posteriors gives
\[
 a_t=1\ \Longrightarrow\ \mu_{t+1}\geq c_t,\qquad
 a_t=0\ \Longrightarrow\ \mu_{t+1}<c_t.
\]
These implications hold simultaneously at all dates outside one null set.
For $k>0$, define
\[
 \tau_t=\frac12+\frac{\mu_{t+1}-1/2}{k},\qquad
 \tau=\frac12+\frac{\mu_\infty-1/2}{k}.
\]
Along such a sample path $\tau_t\to\tau$ and the realized action obeys
$a_t=\1\{z_t\leq\tau_t\}$. The use of $\mu_{t+1}$ here is a pathwise
identity after observing the action, not a claim that the acting agent knows
this posterior in advance.

Suppose $\tau\in(0,1)$. For any sufficiently small $\varepsilon>0$, choose
$T$ so that for $t\geq T$, $|\tau_t-\tau|<\varepsilon/2$ and
$\gamma_t<\varepsilon/2$. If $z_t<\tau-\varepsilon$, the action is 1 and
popularity rises until it enters the $\varepsilon$ neighborhood of $\tau$.
It must enter in finite time: otherwise
$1-z_{t+1}=(1-\gamma_t)(1-z_t)$ would make $z_t\to1$, a contradiction.
From above $\tau+\varepsilon$, the analogous argument using action 0 gives
entry from above. After entry the neighborhood is invariant. In particular,
a downward step can start only above $\tau-\varepsilon/2$ and has size below
$\varepsilon/2$; the upward case is symmetric. Hence $z_t\to\tau$.

If $\tau\leq0$, then eventually $\tau_t<\varepsilon/2$. Above
$\varepsilon$, all actions are 0 until entry into $[0,\varepsilon]$; entry is
forced by the nonsummable gains. Subsequent upward steps can start only below
$\varepsilon/2$ and remain below $\varepsilon$. Thus $z_t\to0$. The case
$\tau\geq1$ is symmetric. This proves $z_t\to\proj_{[0,1]}\tau$.
The cutoff limit follows by continuity. For $k=0$, it is constant at $1/2$;
no unconditional claim about index convergence at $k=0$ is needed.
\end{proof}

\subsection{Proof of Theorem~\ref{thm:gaussian-power}}

\begin{proof}
\emph{Attainment and continuity.} Write $\Phi_N$ and $\phi$ for the standard
normal distribution function and density. For $\rho>0$ the private posterior
is
\[
 Q=\operatorname{logistic}\left(\operatorname{logit}\mu+
          \rho(\theta-1/2)+\sqrt\rho Z\right),\qquad Z\sim N(0,1),
\]
with the null experiment $Q=\mu$ at $\rho=0$. The posterior laws are jointly
weakly continuous in $(\mu,\rho)$ for interior $\mu$, including at
$\rho=0$. At endpoint beliefs this follows from Bayes plausibility: for
example, $\E Q=\mu\to0$ implies $Q\to0$ in probability, uniformly over
bounded precisions. For every experiment and cutoff,
$D_G(\mu,c;\rho)\leq2\mu(1-\mu)\leq1/2$.
Any $\rho>R=(1/(2\kappa))^{1/\beta}$ therefore has base cost above maximal
value. Restricting the menu to $[0,R]$ loses no optimal choice. The cost is
continuous on this compact menu, so Assumption~\ref{ass:continuous-menu}
holds.

\emph{Positive value at alignment.} Put
$g(\rho)=2\Phi_N(\sqrt\rho/2)-1$. Direct evaluation at the aligned
signal threshold $s=1/2$ gives
\[
 D_G(\mu,\mu;\rho)=2\mu(1-\mu)g(\rho),\qquad
 g(\rho)=\phi(0)\sqrt\rho+O(\rho^{3/2}).
\]
If $\beta>1/2$, this exceeds $\kappa\rho^\beta$ for sufficiently small
positive $\rho$ at each interior belief. Hence for $k\geq1$ the criterion
of Theorem~\ref{thm:continued-learning} holds. Full support of the induced
actions at every positive-precision purchase rules out finite purchases on a
path with complete learning.

\emph{An intrinsic obstruction when $\beta\leq1/2$.} For these exponents,
\[
 M_\beta=\sup_{\rho>0}\frac{g(\rho)}{\rho^\beta}<\infty.
\]
Indeed, the ratio is bounded near zero by the displayed expansion and tends
to zero at infinity. At $\beta=1/2$, $M_{1/2}=\phi(0)$, since
$g(\rho)=2\int_0^{\sqrt\rho/2}\phi(v)\,dv\leq\phi(0)\sqrt\rho$.
Consequently, if $2\mu(1-\mu)M_\beta\leq\kappa$, no precision can have
positive net value even at alignment. There are therefore absorbing
neighborhoods of both endpoint beliefs for every $k$.

\emph{An obstruction from insufficient congestion.} Fix $k<1$ and
$\ell=(1-k)/2\in(0,1/2]$. At a small belief $\mu<\ell$, the best feasible
cutoff is $\ell$. Any profitable positive precision would have to satisfy
\[
 \rho<R(\mu):=(2\mu/\kappa)^{1/\beta},
\]
because $D_G\leq2\mu$. Put $L(\mu)=\operatorname{logit}\ell-
\operatorname{logit}\mu$. For sufficiently small $\mu$, $L(\mu)\geq1$ and
$R(\mu)\leq\min\{1,1/(8\beta)\}$. At cutoff $\ell$, the signal threshold
is $s_*=1/2+L(\mu)/\rho$. Let
$A_i=\Prb(s\geq s_*\mid\theta=i)$. The gross value satisfies
\[
 D_G(\mu,\ell;\rho)
 =2[(1-\ell)\mu A_1-\ell(1-\mu)A_0]
 \leq2\mu A_1.
\]
For $0<\rho\leq R(\mu)$, the standardized threshold under state 1 is at
least $1/(2\sqrt\rho)$, so $A_1\leq\exp[-1/(8\rho)]$. The function
$\rho^{-\beta}\exp[-1/(8\rho)]$ is increasing for
$\rho\leq1/(8\beta)$. Therefore
\[
 \frac{D_G(\mu,\ell;\rho)}{\kappa\rho^\beta}
 \leq\frac{2\mu}{\kappa}R(\mu)^{-\beta}
           e^{-1/(8R(\mu))}
 =e^{-1/(8R(\mu))}<1.
\]
For $\rho>R(\mu)$, its cost already exceeds $2\mu$.
There is no profitable precision at sufficiently small $\mu$, for any
$\beta>0$. Symmetry gives an absorbing neighborhood near 1.

\emph{Interior limits in the failure regions.} In either case just proved,
there are absorbing neighborhoods of 0 and 1, uniform over all feasible
popularity values and nonnegative entry fees. Every finite history of
positive-precision Gaussian threshold actions has positive likelihood in both
states, so no finite posterior equals an endpoint. A path converging to an
endpoint would eventually enter its absorbing neighborhood at an interior
belief, where all subsequent experiments would be null and the posterior
would remain fixed. This is impossible. Thus
$\mu_\infty\in(0,1)$ almost surely whenever $k<1$ or $\beta\leq1/2$.

\emph{An endogenous lower bound on purchased precision.} Suppose now
$k<1$ and $\beta>1/2$. The absorbing neighborhoods provide an
$\varepsilon>0$ such that any purchase occurs at
$\mu\in[\varepsilon,1-\varepsilon]$. Cutoffs belong to the compact interval
$C_k\subset(0,1)$. Suppose, contrary to the claim, that a sequence of
optimally purchased precisions $\rho_n$ tends to zero, with associated
$(\mu_n,c_n)\to(\mu,c)$ after taking a subsequence.

If $\mu=c$, the aligned base net value $W(\mu,\mu)$ is strictly positive.
Continuity makes $W(\mu_n,c_n)$ bounded below by a positive constant. A
purchased precision maximizes base net value, so
$W(\mu_n,c_n)=D_G(\mu_n,c_n;\rho_n)-\kappa\rho_n^\beta$.
Joint continuity at the null experiment makes this expression tend to zero,
a contradiction.

If $\mu\ne c$, the log-odds distance
$|\operatorname{logit}c_n-\operatorname{logit}\mu_n|$ is bounded below by a
positive constant $b$. Gaussian tails, by the same threshold argument as
above (and its symmetric version), give
\[
 D_G(\mu_n,c_n;\rho_n)\leq2e^{-b^2/(8\rho_n)}
\]
for large $n$. Dividing by $\kappa\rho_n^\beta$ gives a ratio tending to
zero, contradicting profitable purchase. This proves a strictly positive
lower bound $\underline\rho$ over all optimally purchased experiments.
If there are no purchases the assertion is vacuous. Otherwise every purchase
has $D_t>\kappa\underline\rho^\beta$. Sum the squared-value budget
\eqref{eq:p3-budget} to obtain the expected-count bound in the theorem.
\end{proof}

\subsection{Proof of Proposition~\ref{prop:gaussian-cycle}}

\begin{proof}
At $\mu_1=c_1=1/2$, the net value is
$\tfrac12[2\Phi_N(\sqrt\rho/2)-1]-\rho$. Its derivative vanishes exactly
when $\sqrt\rho=\phi(\sqrt\rho/2)/4$. The left side strictly increases
and the right side strictly decreases, so the solution is unique and
maximizes net value. Small precisions have strictly positive value, so it is
purchased. Since $\phi(0)<0.4$, $\sqrt{\rho_*}<0.1$. Therefore
\[
 0<r_*-1/2=\int_0^{\sqrt{\rho_*}/2}\phi(v)\,dv<0.02.
\]
The two actions induce posteriors $r_*$ and $1-r_*$ and send popularity to
$z_2=0.9$ and $0.1$, respectively.

We show that no experiment is profitable when
$\mu\in[0.48,0.52]$ and $z\in[0,0.2]\cup[0.8,1]$.
Here $c=z$ and $a=|\mu-c|\geq1/4$. For any mean-$\mu$ posterior $Q$, the
inequality $(y-a)_+\leq y^2/(4a)$ and its reflected version imply
\[
 D(\mu,c)\leq\frac{\E(Q-\mu)^2}{2a}.
\]
Let $L=\rho(\theta-1/2)+\sqrt\rho Z$ be the Gaussian log likelihood ratio.
The logistic map has derivative at most $1/4$, so
\[
 |Q-\mu|\leq|L|/4,\qquad
 \E(Q-\mu)^2\leq\frac{\rho+\rho^2/4}{16}.
\]
For $0<\rho\leq1/2$, it follows that
\[
 D_G(\mu,c;\rho)\leq\frac{\rho+\rho^2/4}{8}
 \leq\frac9{64}\rho<\rho=C(\rho).
\]
For $\rho>1/2$, the universal bound $D_G\leq1/2$ also makes purchase
unprofitable. Thus null is uniquely optimal on this set.

A null action at $z\leq0.2$ is 1 and maps $z$ to $0.8+0.2z\in[0.8,0.84]$.
At $z\geq0.8$ it is 0 and maps $z$ to $0.2z\in[0.16,0.2]$.
The union is invariant, and beliefs remain at their first-action value.
Actions alternate forever. The two-step map is a contraction with fixed
cycle points $1/6$ and $5/6$. Alternation gives a correct-action frequency
of $1/2$ under either state. Finally, the aligned small-precision expansion
with $\beta=1$ gives $G_1(\mu)>0$ at every interior belief.
\end{proof}

\section{Binary values and terminal outcomes}\label{app:binary-proofs}

\subsection{Value of the binary signal}

\begin{proposition}\label{prop:binary-value}
Let $c=c_k(p)$.  The gross value of purchasing the binary signal relative to
acting on the public belief is
\begin{equation}\label{eq:binary-value-absolute}
    D_r(\mu,c)
    =\E\bigl[\lvert Q-c\rvert\bigr]-\lvert\mu-c\rvert,
\end{equation}
where $Q=q_y(\mu)$ after signal realization $y$.  Equivalently,
\begin{equation}\label{eq:binary-value-piecewise}
D_r(\mu,c)=
\begin{cases}
0,
    & c\leq q_0(\mu),\\[3pt]
2\bigl[\pi_0(\mu)c-\mu(1-r)\bigr],
    & q_0(\mu)<c\leq\mu,\\[3pt]
2\bigl[\mu r-\pi_1(\mu)c\bigr],
    & \mu\leq c<q_1(\mu),\\[3pt]
0,
    & c\geq q_1(\mu).
\end{cases}
\end{equation}
For fixed $\mu$, the value is weakly increasing on $(-\infty,\mu]$, weakly
decreasing on $[\mu,\infty)$, and maximized at $c=\mu$, where
\begin{equation}\label{eq:aligned-value}
    D_r(\mu,\mu)=2(2r-1)\mu(1-\mu).
\end{equation}
The agent purchases the signal if and only if $D_r(\mu,c)>F$.
\end{proposition}

\begin{corollary}\label{cor:intrinsic}
Some cutoff induces acquisition if and only if $\mu\in(a^*,1-a^*)$, with
$a^*$ defined in \eqref{eq:astar}.
\end{corollary}
\begin{proof}
By \eqref{eq:max-payoff}, expected gross payoff after purchasing the
signal is
\[
    \frac{1-k}{2}+\E\lvert Q-c\rvert,
\]
whereas acting without the signal yields
$(1-k)/2+\lvert\mu-c\rvert$.  Their difference is
\eqref{eq:binary-value-absolute}.

Bayes plausibility and Bayes' rule give
\begin{equation}\label{eq:binary-bayes-identities}
    \pi_1q_1+\pi_0q_0=\mu,\qquad
    \pi_1q_1=\mu r,\qquad
    \pi_0q_0=\mu(1-r).
\end{equation}
If $c\leq q_0$, both posterior realizations lie above the cutoff, so
$\E\lvert Q-c\rvert=\mu-c=\lvert\mu-c\rvert$ and the value is zero.
The same argument applies when $c\geq q_1$.  If $q_0<c\leq\mu$, direct
expansion and \eqref{eq:binary-bayes-identities} give
\[
    D_r(\mu,c)
    =2\pi_0(c-q_0)
    =2\bigl[\pi_0(\mu)c-\mu(1-r)\bigr].
\]
If $\mu\leq c<q_1$, the symmetric calculation gives
\[
    D_r(\mu,c)
    =2\pi_1(q_1-c)
    =2\bigl[\mu r-\pi_1(\mu)c\bigr].
\]
This proves \eqref{eq:binary-value-piecewise}.  The slopes on the two interior
pieces are $2\pi_0>0$ and $-2\pi_1<0$, respectively, so the maximum occurs at
$c=\mu$.  Substitution yields
\[
    D_r(\mu,\mu)=2(2r-1)\mu(1-\mu).
\]
The tie-breaking convention implies purchase exactly when this gross value at
the prevailing cutoff is strictly greater than $F$.

For Corollary~\ref{cor:intrinsic}, write $d=2r-1$.  Some cutoff makes purchase
profitable if and only if
\[
    2d\mu(1-\mu)>F.
\]
The two roots of the associated equality are $a^*$ and $1-a^*$ as defined in
\eqref{eq:astar}.  Assumption~\ref{ass:binary} ensures that the roots are real
and lie strictly on either side of $1/2$.  The quadratic is above $F$ exactly
between them.
\end{proof}

\subsection{Proof of Proposition~\ref{prop:binary-restart}}

\begin{proof}
Proposition~\ref{prop:binary-value} implies that, at a fixed belief, signal
value is maximized by the feasible cutoff closest to $\mu$.  Since
$c_k([0,1])=C_k$,
\[
    \overline D_k(\mu)
    =D_r\bigl(\mu,\proj_{C_k}\mu\bigr).
\]
The model is symmetric around $1/2$, so it suffices to consider
$\mu\leq1/2$.  The closest feasible cutoff is
$\max\{\mu,\ell_k\}$.

Suppose first that $\ell_k\leq a^*$.  If $\mu>a^*$, then
$\ell_k<\mu$ and the aligned cutoff is feasible, so the best value exceeds
$F$.  If $\mu\leq a^*$, even the aligned value is at most $F$.  The lower
restart boundary is therefore $a^*$.

Now suppose $\ell_k>a^*$.  This case implies $\ell_k\in(0,1/2]$.  For
$\mu\geq\ell_k$, alignment is feasible and the value exceeds $F$.  For
$\mu<\ell_k$, the best cutoff is $\ell_k$ and
\eqref{eq:binary-value-piecewise} gives the best value as
$\max\{0,B_k(\mu)\}$, where
\[
    B_k(\mu)
    =2\left[
      \mu r-\bigl((1-r)+(2r-1)\mu\bigr)\ell_k
    \right].
\]
This expression is strictly increasing because
$r-(2r-1)\ell_k>0$.  At $\mu=\ell_k$ it equals
$2(2r-1)\ell_k(1-\ell_k)>F$, whereas it is below $F$ for sufficiently small
$\mu$.  Solving $B_k(\mu)=F$ gives
\[
    \mu
    =\frac{F/2+(1-r)\ell_k}
    {r-(2r-1)\ell_k}
    =a_k.
\]
Thus the lower restart region is $(a_k,1/2]$, and symmetry gives
$(a_k,1-a_k)$.

If $\mu$ lies outside this interval, the maximal feasible value is at most
$F$, so no popularity history induces acquisition.  If $\mu$ is inside it,
Lemma~\ref{lem:ongoing-tracking} gives
$c_k(p_t)\to\widehat c_k(\mu)$.  Continuity of $D_r$ and the strict inequality
$\overline D_k(\mu)>F$ imply that acquisition becomes strictly optimal after
finitely many no-acquisition actions.

On the second branch of \eqref{eq:ak}, treating $a_k$ as a function of
$\ell_k$ gives
\[
    \frac{\partial a_k}{\partial\ell_k}
    =
    \frac{r(1-r)+(2r-1)F/2}
    {\bigl[r-(2r-1)\ell_k\bigr]^2}>0.
\]
Since $\ell_k=(1-k)/2$ decreases in $k$, $a_k$ decreases until
$\ell_k=a^*$, after which it equals $a^*$.  The equality
$\ell_k\leq a^*$ is equivalent to \eqref{eq:k-saturation}.
\end{proof}

\subsection{Proof of Theorem~\ref{thm:binary-random-walk}}

\begin{proof}
Whenever the signal is purchased,
$D_r(\mu,c)>F>0$.  Equation \eqref{eq:binary-value-piecewise} then implies
$q_0(\mu)<c<q_1(\mu)$.  The cutoff rule maps signal $0$ into action $0$ and
signal $1$ into action $1$, so the public action reveals the signal.

Let $L=\log(\mu/(1-\mu))$ denote public log odds.  Bayes' rule gives
\[
    L'=
    \begin{cases}
        L+\lambda, & y=1,\\
        L-\lambda, & y=0.
    \end{cases}
\]
At no-acquisition dates the action is deterministic and $L$ does not move.
Proposition~\ref{prop:binary-restart} guarantees another information date
whenever the current belief lies in $(a_k,1-a_k)$ and guarantees no further
purchase once it lies outside.  Starting from $L=0$, the embedded process
therefore stops at the first integer multiple of $\lambda$ whose magnitude is
at least $\log((1-a_k)/a_k)$.  This is $\pm H_k\lambda$ with $H_k$ defined in
\eqref{eq:Hk}.

Conditional on $\theta=1$, an upward step occurs with probability $r$ and a
downward step with probability $1-r$.  For a biased nearest-neighbor random
walk starting at zero with barriers $-H$ and $H$, solving the standard
two-boundary difference equations gives
\[
    \Prb(\text{hit }-H\text{ before }H)
    =\frac{x^H}{1+x^H}
\]
and
\[
    \E[T]
    =\frac{H(1-x^H)}{(2r-1)(1+x^H)}.
\]
Taking $H=H_k$ proves \eqref{eq:error} and
\eqref{eq:expected-purchases}.  State symmetry gives the same expressions
conditional on $\theta=0$.

Proposition~\ref{prop:binary-restart} shows that $a_k$ is nonincreasing in
$k$, so $H_k$ is nondecreasing.  Moving absorbing barriers weakly farther from
the starting point lowers the probability of hitting the wrong barrier and
raises expected absorption time.  Once $a_k=a^*$, all expressions are
constant in $k$.
\end{proof}

\subsection{Proof of Proposition~\ref{prop:accuracy}}

\begin{proof}
At the upper stopping point, public log odds are $H_k\lambda$, so the public
belief is
\[
    \frac{e^{H_k\lambda}}{1+e^{H_k\lambda}}
    =\frac{1}{1+x^{H_k}}=u_k.
\]
The gambler's-ruin probability in Theorem~\ref{thm:binary-random-walk} also
implies that, conditional on $\theta=1$, the upper point is reached with
probability $u_k$ and the lower point with probability $1-u_k$.

After information acquisition stops at the upper belief, the limiting
popularity for $k>0$ from Lemma~\ref{lem:ongoing-tracking} is $z_k$ in
\eqref{eq:zk}; for $k=0$, every subsequent action is 1. By symmetry, the limiting frequency of action $1$ at the lower
belief is $1-z_k$.  Since information acquisition stops after finitely many
dates, these popularity limits also establish almost-sure existence of the
limiting correct-action frequency.  Conditional on $\theta=1$, its expectation
is
\[
    u_kz_k+(1-u_k)(1-z_k).
\]
State symmetry gives the same ex ante expression.

For $k>0$, the target in \eqref{eq:zk} reaches the upper boundary if and only
if
\[
    \frac12+\frac{u_k-1/2}{k}\geq1
    \quad\Longleftrightarrow\quad
    k\leq2u_k-1.
\]
In that case $z_k=1$ and $A_k=u_k$.  Otherwise,
$z_k=1/2+(2u_k-1)/(2k)$.  Substitution into \eqref{eq:Ak} yields
\[
    A_k=\frac12+\frac{(2u_k-1)^2}{2k}.
\]
On a plateau of $H_k$, $u_k$ is constant, so the latter expression decreases
to $1/2$ as $k$ rises.
\end{proof}

\subsection{Proof of Corollary~\ref{cor:three-regimes}}

\begin{proof}
For an integer $h\geq1$, let
\[
    b_h=\frac{x^h}{1+x^h}=1-\frac{1}{1+x^h}.
\]
If $H_k=h$, the definition of the ceiling in \eqref{eq:Hk} implies
$b_h\leq a_k$.  Suppose first that $k<k_R$.  Then
$\ell_k>a^*$, so the second branch of \eqref{eq:ak} applies.  At
$\mu=\ell_k$, the feasible cutoff is aligned and its signal value is strictly
above $F$ because $\ell_k>a^*$.  Since the relevant value branch is strictly
increasing in $\mu$, its root satisfies $a_k<\ell_k$.  With $h=H_k$, we
therefore have
\[
    b_h\leq a_k<\ell_k=\frac{1-k}{2},
\]
and hence
\[
    k<1-2b_h=2u_k-1.
\]
Proposition~\ref{prop:accuracy} then gives $A_k=u_k$, and all actions after
shutdown follow the sign of the terminal belief.  Proposition~\ref{prop:binary-restart}
and Theorem~\ref{thm:binary-random-walk} establish the remaining weak
monotonicities in the first regime.

For $k\geq k_R$, we have $\ell_k\leq a^*$, so $a_k=a^*$ and
$H_k=H^*$.  Applying the ceiling argument once more gives
\[
    1-u^*=b_{H^*}\leq a^*.
\]
Thus
\[
    k_A=2u^*-1=1-2b_{H^*}
    \geq1-2a^*=k_R.
\]
Assumption~\ref{ass:binary} gives $0<k_R$; the strict interiority of
$b_{H^*}$ gives $k_A<1$.  Proposition~\ref{prop:accuracy} yields
$A_k=u^*$ for $k\leq k_A$ and the second line of
\eqref{eq:Ak-piecewise} for $k>k_A$.
The latter expression is strictly decreasing in $k$ and converges to $1/2$.
For $0<k<2u_k-1$, the targets at the upper and lower terminal beliefs lie
strictly above one and below zero, respectively, so every post-shutdown action
follows the terminal belief.  The same conclusion is immediate at $k=0$.
At $k=k_A$, the targets are one and zero.  At the lower terminal belief with
popularity zero, Assumption~\ref{ass:ties} produces one action $1$; popularity
then becomes positive and every subsequent action is $0$.  This transient has
no effect on the limiting frequency.  When $k>k_A$, the upper target lies
strictly inside $(1/2,1)$, and popularity targeting assigns a positive limiting
frequency to action $0$ even at the upper terminal belief.  This proves the
terminal-action claims.
\end{proof}

\subsection{Proof of Proposition~\ref{prop:declining-binary}}

\begin{proof}
Treat the menu as the null and the fixed binary experiment, both with zero
base cost. It is compact and its posterior laws are continuous in $\mu$.
For $k\geq1$, maximal attainable value is
$G_k(\mu)=2(2r-1)\mu(1-\mu)>0$ at every interior belief. Theorem~\ref{thm:continued-learning}
gives complete learning. A finite number of full-support signals leaves an
interior posterior, which thereafter stays fixed if there are no further
purchases. Thus purchases must be infinite almost surely.

For $k<1$, put $\ell=(1-k)/2>0$. By symmetry take $\mu\leq1/2$.
When $\mu\geq\ell$, alignment is feasible and value is positive. When
$\mu<\ell$, the best attainable cutoff is $\ell$ and value is
\[
 \max\{0,2[(r-d\ell)\mu-(1-r)\ell]\}.
\]
It is positive precisely when $\mu>\alpha_k$.
Thus $G_k$ is zero outside $(\alpha_k,1-\alpha_k)$, where the null experiment
is selected at every date. At a belief inside this interval, a hypothetical
infinite null spell would have a cutoff converging to its attainable target.
Positive value there, continuity, and $F_t\to0$ make purchase strictly
profitable after finitely many dates, a contradiction.

Whenever a binary signal is purchased, its positive gross value implies
$q_0(\mu)<c_t<q_1(\mu)$; the action reveals the signal. Conditional on the
state, signals drawn at these predictable purchase dates are independent with
accuracy $r$. Every interior node of the finite purchase-clock interval is
followed in finite calendar time by another purchase, while the boundary
nodes are absorbing. The open interval gives the stated ceiling rule,
including at integer values. The finite nearest-neighbor chain is absorbed
almost surely. Solving its hitting-probability recurrence with boundaries
$\pm H$ gives wrong-boundary probability $x^H/(1+x^H)$. Solving the
expected-hitting-time recurrence gives $H(1-x^H)/(d(1+x^H))$.
These expressions do not depend on the calendar waiting times or fee path.
Symmetry gives the same formulas in the other state.

\end{proof}

\section{Supplementary binary calculations}\label{app:binary-extra}

\subsection{Calendar delays}

For any $\mu\in(a^*,1-a^*)$, define
\begin{align}
    c_L(\mu)
    &=\frac{F/2+\mu(1-r)}{\pi_0(\mu)}, &
    c_U(\mu)
    &=\frac{\mu r-F/2}{\pi_1(\mu)}. \label{eq:c-purchase-bounds}
\end{align}
Then
\[
    c_L(\mu)<\mu<c_U(\mu),
\]
and the signal is purchased exactly when $c\in(c_L(\mu),c_U(\mu))$.
For $k>0$, map these cutoffs into popularity thresholds:
\begin{equation}\label{eq:p-purchase-bounds}
    p_L(\mu,k)=\frac12+\frac{c_L(\mu)-1/2}{k},
    \qquad
    p_U(\mu,k)=\frac12+\frac{c_U(\mu)-1/2}{k}.
\end{equation}

\begin{proposition}\label{prop:waiting}
Fix $k>0$ and a belief in the restart region.  Every no-information spell ends
after finitely many actions.  Consider a sequence of such spells beginning
after $n$ actions, of which $m_n$ are action $1$, and let $s_n$ be the number
of additional no-acquisition actions.  Whenever the corresponding starting
histories are feasible,
\begin{align}
    \frac{m_n}{n}\to\bar p<p_L(\mu,k)
    &\quad\Longrightarrow\quad
    \frac{s_n}{n}\to
    \frac{p_L(\mu,k)-\bar p}{1-p_L(\mu,k)},
    \label{eq:wait-up-limit}\\
    \frac{m_n}{n}\to\bar p>p_U(\mu,k)
    &\quad\Longrightarrow\quad
    \frac{s_n}{n}\to
    \frac{\bar p-p_U(\mu,k)}{p_U(\mu,k)}.
    \label{eq:wait-down-limit}
\end{align}
The proof gives the exact finite-history formulas.
\end{proposition}

\begin{proof}
At a belief in the intrinsic interval, solving $D_r(\mu,c)>F$ on the two
linear branches of \eqref{eq:binary-value-piecewise} gives
\[
    c>c_L(\mu)\quad\text{for }c\leq\mu,
    \qquad
    c<c_U(\mu)\quad\text{for }c\geq\mu.
\]
The aligned value exceeds $F$, so $c_L(\mu)<\mu<c_U(\mu)$.  Since
$c_k(p)$ is strictly increasing in $p$ when $k>0$, purchase is equivalent to
$p\in(p_L,p_U)$.

Finite entry follows directly from Proposition~\ref{prop:binary-restart}.
Consider first a sequence with $m_n/n\to\bar p<p_L$.  For all sufficiently
large $n$, the starting popularity is below $p_L$ and
\[
    \frac{1}{n+1}<p_U-p_L.
\]
Until popularity first exceeds $p_L$, acquisition is not profitable and every
action is $1$.  The exact number of interim actions is
\begin{equation}\label{eq:wait-up-exact}
    s_n=
    \left\lfloor
    \frac{p_L n-m_n}{1-p_L}
    \right\rfloor+1,
\end{equation}
because this is the smallest positive integer for which
$(m_n+s_n)/(n+s_n)>p_L$.  The one-period change is smaller than
$p_U-p_L$, so the crossing cannot jump over $p_U$ and the next agent
purchases.  Dividing \eqref{eq:wait-up-exact} by $n$ yields
\eqref{eq:wait-up-limit}.

If $m_n/n\to\bar p>p_U$, every interim action is $0$ for all sufficiently large
$n$.  The symmetric exact formula is
\begin{equation}\label{eq:wait-down-exact}
    s_n=
    \left\lfloor
    \frac{m_n}{p_U}-n
    \right\rfloor+1.
\end{equation}
The same update-size argument puts the crossing inside $(p_L,p_U)$, and
division by $n$ gives \eqref{eq:wait-down-limit}.  In both cases the floor
error divided by $n$ vanishes.
\end{proof}

\subsection{A vanishing fee across stationary economies}

\begin{corollary}\label{cor:vanishing-cost}
As $F\downarrow0$:
\begin{enumerate}[label=(\roman*)]
    \item if $k<1$, then
    \begin{equation}\label{eq:ak-limit}
        a_k\longrightarrow
        \frac{(1-r)(1-k)/2}
        {r-(2r-1)(1-k)/2}>0.
    \end{equation}
    Thus $H_k$ remains bounded and terminal-belief error remains bounded away
    from zero;
    \item if $k\geq1$, then $a_k=a^*\to0$, $H_k\to\infty$, and terminal-belief
    error converges to zero.  Moreover,
    \[
        H_k=\frac{\log(1/F)}{\lambda}+O(1),
        \qquad
        \E[N_I\mid\theta]
        =\frac{\log(1/F)}{(2r-1)\lambda}+O(1).
    \]
\end{enumerate}
Thus the threshold $k=1$ from the costless congested-learning benchmark
\citep{EysterEtAl2014} reappears in the vanishing-fixed-cost limit.  For every
$F>0$, however, the total number of purchases is finite.
\end{corollary}

This limit compares stationary economies with different positive fees. It is
not the calendar-time fee sequence of Proposition~\ref{prop:declining-binary}.
An automatically supplied free signal can change an action at indifference
even when optional acquisition would be rejected at zero value.

\begin{proof}
Fix $k<1$.  Then $\ell_k=(1-k)/2>0$, whereas $a^*\to0$.  For all sufficiently
small $F$, the second branch of \eqref{eq:ak} applies, and taking the limit
gives \eqref{eq:ak-limit}.  The limit is strictly positive, so $H_k$ remains
bounded and the error in \eqref{eq:error} remains bounded away from zero.

If $k\geq1$, then $\ell_k\leq0<a^*$ for every $F>0$, so $a_k=a^*$.  A Taylor
expansion gives
\[
    a^*
    =\frac{F}{2(2r-1)}+O(F^2).
\]
Consequently,
\[
    \log\frac{1-a^*}{a^*}
    =\log\frac{1}{F}+O(1),
\]
which proves the expansion for $H_k$.  Since $x\in(0,1)$ and $H_k\to\infty$,
$x^{H_k}\to0$.  Substitution into \eqref{eq:error} and
\eqref{eq:expected-purchases} proves the remaining claims.
\end{proof}


\begin{thebibliography}{99}

\bibitem[Acemoglu et al.(2011)Acemoglu, Dahleh, Lobel, and Ozdaglar]
{AcemogluEtAl2011}
Acemoglu, Daron, Munther A. Dahleh, Ilan Lobel, and Asuman Ozdaglar (2011).
\newblock Bayesian learning in social networks.
\newblock \emph{Review of Economic Studies} 78(4), 1201--1236.
\newblock \href{https://doi.org/10.1093/restud/rdr004}{doi:10.1093/restud/rdr004}.

\bibitem[Ali and Kartik(2012)]{AliKartik2012}
Ali, S. Nageeb, and Navin Kartik (2012).
\newblock Herding with collective preferences.
\newblock \emph{Economic Theory} 51(3), 601--626.
\newblock \href{https://doi.org/10.1007/s00199-011-0609-7}
{https://doi.org/10.1007/s00199-011-0609-7}.

\bibitem[Ali(2018)]{Ali2018}
Ali, S. Nageeb (2018).
\newblock Herding with costly information.
\newblock \emph{Journal of Economic Theory} 175, 713--729.
\newblock \href{https://doi.org/10.1016/j.jet.2018.02.009}
{https://doi.org/10.1016/j.jet.2018.02.009}.

\bibitem[Arieli(2017)]{Arieli2017}
Arieli, Itai (2017).
\newblock Payoff externalities and social learning.
\newblock \emph{Games and Economic Behavior} 104, 392--410.
\newblock \href{https://doi.org/10.1016/j.geb.2017.05.005}
{https://doi.org/10.1016/j.geb.2017.05.005}.

\bibitem[Bernhardt et al.(2006)Bernhardt, Campello, and Kutsoati]
{BernhardtEtAl2006}
Bernhardt, Dan, Murillo Campello, and Edward Kutsoati (2006).
\newblock Who herds?
\newblock \emph{Journal of Financial Economics} 80(3), 657--675.
\newblock \href{https://doi.org/10.1016/j.jfineco.2005.07.006}
{https://doi.org/10.1016/j.jfineco.2005.07.006}.

\bibitem[Besbes and Scarsini(2018)]{BesbesScarsini2018}
Besbes, Omar, and Marco Scarsini (2018).
\newblock On information distortions in online ratings.
\newblock \emph{Operations Research} 66(3), 597--610.
\newblock \href{https://doi.org/10.1287/opre.2017.1676}{doi:10.1287/opre.2017.1676}.


\bibitem[Bikhchandani et al.(2024)Bikhchandani, Hirshleifer, Tamuz, and Welch]
{BikhchandaniEtAl2024}
Bikhchandani, Sushil, David Hirshleifer, Omer Tamuz, and Ivo Welch (2024).
\newblock Information cascades and social learning.
\newblock \emph{Journal of Economic Literature} 62(3), 1040--1093.
\newblock \href{https://doi.org/10.1257/jel.20241472}
{https://doi.org/10.1257/jel.20241472}.

\bibitem[Bobkova and Mass(2022)]{BobkovaMass2022}
Bobkova, Nina, and Helene Mass (2022).
\newblock Two-dimensional information acquisition in social learning.
\newblock \emph{Journal of Economic Theory} 202, 105451.
\newblock \href{https://doi.org/10.1016/j.jet.2022.105451}
{https://doi.org/10.1016/j.jet.2022.105451}.

\bibitem[Burguet and Vives(2000)]{BurguetVives2000}
Burguet, Roberto, and Xavier Vives (2000).
\newblock Social learning and costly information acquisition.
\newblock \emph{Economic Theory} 15(1), 185--205.
\newblock \href{https://doi.org/10.1007/s001990050006}
{https://doi.org/10.1007/s001990050006}.

\bibitem[Eyster et al.(2014)Eyster, Galeotti, Kartik, and Rabin]
{EysterEtAl2014}
Eyster, Erik, Andrea Galeotti, Navin Kartik, and Matthew Rabin (2014).
\newblock Congested observational learning.
\newblock \emph{Games and Economic Behavior} 87, 519--538.
\newblock \href{https://doi.org/10.1016/j.geb.2014.06.006}
{https://doi.org/10.1016/j.geb.2014.06.006}.

\bibitem[Heinsalu(2019)]{Heinsalu2019}
Heinsalu, Sander (2019).
\newblock Herding driven by the desire to differ.
\newblock arXiv:1904.00454, March 2019.
\newblock \href{https://arxiv.org/abs/1904.00454}
{https://arxiv.org/abs/1904.00454}.

\bibitem[Hendricks et al.(2012)Hendricks, Sorensen, and Wiseman]
{HendricksEtAl2012}
Hendricks, Kenneth, Alan Sorensen, and Thomas Wiseman (2012).
\newblock Observational learning and demand for search goods.
\newblock \emph{American Economic Journal: Microeconomics} 4(1), 1--31.
\newblock \href{https://doi.org/10.1257/mic.4.1.1}
{https://doi.org/10.1257/mic.4.1.1}.

\bibitem[Kartik et al.(2024)Kartik, Lee, Liu, and Rappoport]
{KartikEtAl2024}
Kartik, Navin, SangMok Lee, Tianhao Liu, and Daniel Rappoport (2024).
\newblock Beyond unbounded beliefs: How preferences and information interplay
in social learning.
\newblock \emph{Econometrica} 92(4), 1033--1062.
\newblock \href{https://doi.org/10.3982/ECTA21470}
{https://doi.org/10.3982/ECTA21470}.

\bibitem[Laster et al.(1999)Laster, Bennett, and Geoum]{LasterEtAl1999}
Laster, David, Paul Bennett, and In Sun Geoum (1999).
\newblock Rational bias in macroeconomic forecasts.
\newblock \emph{Quarterly Journal of Economics} 114(1), 293--318.
\newblock \href{https://doi.org/10.1162/003355399555918}
{https://doi.org/10.1162/003355399555918}.

\bibitem[Mueller-Frank and Pai(2016)]{MuellerFrankPai2016}
Mueller-Frank, Manuel, and Mallesh M. Pai (2016).
\newblock Social learning with costly search.
\newblock \emph{American Economic Journal: Microeconomics} 8(1), 83--109.
\newblock \href{https://doi.org/10.1257/mic.20130253}
{https://doi.org/10.1257/mic.20130253}.

\bibitem[Ottaviani and S{\o}rensen(2006)]{OttavianiSorensen2006}
Ottaviani, Marco, and Peter Norman S{\o}rensen (2006).
\newblock The strategy of professional forecasting.
\newblock \emph{Journal of Financial Economics} 81(2), 441--466.
\newblock \href{https://doi.org/10.1016/j.jfineco.2005.06.003}
{https://doi.org/10.1016/j.jfineco.2005.06.003}.

\bibitem[Smith et al.(2021)Smith, S{\o}rensen, and Tian]
{SmithSorensenTian2021}
Smith, Lones, Peter Norman S{\o}rensen, and Jianrong Tian (2021).
\newblock Informational herding, optimal experimentation, and contrarianism.
\newblock \emph{Review of Economic Studies} 88(5), 2527--2554.
\newblock \href{https://doi.org/10.1093/restud/rdab001}
{https://doi.org/10.1093/restud/rdab001}.

\bibitem[Song(2025)]{Song2025}
Song, Yangbo (2025).
\newblock Social learning among opinion leaders.
\newblock \emph{Games and Economic Behavior} 153, 451--473.
\newblock \href{https://doi.org/10.1016/j.geb.2025.07.011}
{https://doi.org/10.1016/j.geb.2025.07.011}.

\bibitem[Yang(2011)]{Yang2011}
Yang, Wan-Ru (2011).
\newblock Herding with costly information and signal extraction.
\newblock \emph{International Review of Economics \& Finance} 20(4), 624--632.
\newblock \href{https://doi.org/10.1016/j.iref.2010.12.004}
{https://doi.org/10.1016/j.iref.2010.12.004}.

\end{thebibliography}
\end{document}